\documentclass[aps,prl,twocolumn,showpacs,reprint,superscriptaddress,longbibliography]{revtex4-1}

\usepackage{amsmath, amsfonts, amssymb, bm}
\usepackage{float}
\usepackage{mathtools}
\usepackage[inline]{enumitem}
\usepackage{graphicx}
\usepackage{xcolor}
\usepackage[breaklinks,hypertexnames=false]{hyperref}
\hypersetup{colorlinks,linkcolor={magenta},citecolor={blue},urlcolor={blue}}
\usepackage[capitalize]{cleveref}
\usepackage{lineno}

\begin{document}
\title{$\varphi$ Josephson junction induced by altermagnetism}

\newcommand{\tianjin}{Department of Physics, Tianjin University, Tianjin 300072, China}

\newcommand{\nagoya}{Department of Applied Physics, Nagoya University, Nagoya 464-8603, Japan}

\author{Bo Lu}
\affiliation{\tianjin}

\author{Kazuki Maeda}
\affiliation{\nagoya}

\author{Hiroyuki Ito}
\affiliation{\nagoya}

\author{Keiji Yada}
\affiliation{\nagoya}

\author{Yukio Tanaka}
\affiliation{\nagoya}

\date{\today}

\begin{abstract}

We study the Josephson effect in a superconductor/altermagnet/superconductor (S/AM/S) junction.
We find anomalous phenomena including $0$-$\pi$ transition as well as multi-nodal current-phase relations. Similar to $d$-wave superconductor, $d$-wave altermagnet can support $\varphi$ junction where free energy minima locate neither $\varphi=0$ nor $\pm \pi$ with double degeneracy. These properties can be tunable by parameters, e.g., the exchange energy, the orientation of crystal axis, the length, and the chemical doping of altermagnet.  These rich features lead to accessible functionality of S/AM/S junction. 

\end{abstract}
\maketitle

\emph{Introduction.---}  Recently, altermagnet (AM)  \cite{LiborSAv,Libor22,Hayami19,Hayami20,landscape22,MazinPNAS,MazinPRX22,Libor011028} has emerged as a new class of magnetic materials distinct from ferro- and antiferromagnet. AM material exhibits spin-polarized Fermi surface resembling ferromagnet but with a collinear compensated magnetic ordering like antiferromagnet. 
AM has been found in various types of materials like metallic ${\mathrm{RuO}}_{2}$ \cite{Ahn19,Libor22}, ${\mathrm{Mn}_{5} \mathrm{Si}_{3}}$  \cite{Helena2021}, semiconducting/insulating ${\mathrm{La}_{2}\mathrm{CuO}_{4}}$ \cite{Moreno16} and ${\mathrm{MnTe}}$ \cite{Lee24,Osumi2024,krempasky2024}, and many more. 

Due to the vanishing net macroscopic magnetization, AM provides a new benefit in combing with superconductors and may have intriguing implications. 
Now, research on the transport properties in junctions consisting of AM and superconductors become a hot topic \cite{Sun23,Papaj23,Beenakker23,Ouassou23,zhang2024,nagae2024,maeda2024p}.  In an AM/$s$-wave superconductor (S) junction, studies show that Andreev reflection is sensitive to both the
crystal orientation and the strength of the spin-splitting field \cite{Sun23,Papaj23}, as compared to ferromagnetic materials which are orientation independent. 
Another remarkable finding is that the Josephson currents through AM also display $0$-$\pi$ oscillations even without any net magnetization \cite{Ouassou23,zhang2024}. Such phenomena are also explained by a phase-shift that depends on the crystalline axis of AM \cite{Beenakker23}. Furthermore, 
in S/AM/spin-triplet superconductor junctions, the $\phi_0$ phase as well as the $0$-$\pi$ transition can be realized as the unique interplay between altermagnetism and spin-triplet Cooper pairs \cite{Cheng24}.  

It can be shown that in a S/AM/S junction, the current-phase relation (CPR) has symmetry $I(\varphi)=-I(-\varphi)$, which excludes the possibility of $\varphi_0$ junction. However, to the best of our knowledge, studies so far reveal only two types of CPR in the S/AM/S junction: $0$- and $\pi$- junctions.   Whether the altermagnetic ordering can generate more exotic CPR, like $\varphi$ junction \cite{Buzdin03}, where the free-energy minimum of the S/AM/S junction locates neither 0 nor $\pi$, can exist or not, remains an open question. 
It is noted that a $\varphi$ junction has a doubly degenerate ground state, which was experimentally observed in the Josephson junction with a current injector \cite{menditto2018evidence}. 
It has been shown that $\varphi$ junctions exhibit interesting physical properties such as non-Fraunhofer interference pattern under an external magnetic field 
\cite{Goldovin,alidoust2013varphi}, half-integer Shapiro steps \cite{Goldovin}, and fractional vortex \cite{mints2001josephson,Goldovin}.  One important route to form $\varphi$ junction is the superposition of multiple $0$- and $\pi$-segments, such as ferromagnetic junctions \cite{heim2013ferromagnetic,goldobin2011josephson,sickinger2012experimental} or d-wave Josephson junctions \cite{Ilichev1999,Lofwander_2001,Testa2005,Tafuri2005,Goldovin}. 
For example, in the Josephson effect with $d$-wave pairings, the current component becomes either positive or negative depending on the injection angle of the quasiparticle \cite{TK95,tanaka961,tanaka971, Yip1995, Barash1996, Kashiwaya_2000}. Thus the total Josephson current as a superposition displays exotic CPR and temperature dependence of maximum Josephson current \cite{tanaka961,tanaka971,Barash1996}. It is quite natural to anticipate that Josephson junction with $d$-wave AM has similar physical properties. 

\begin{figure}[t]
	\begin{center}
		\includegraphics[width=87mm]{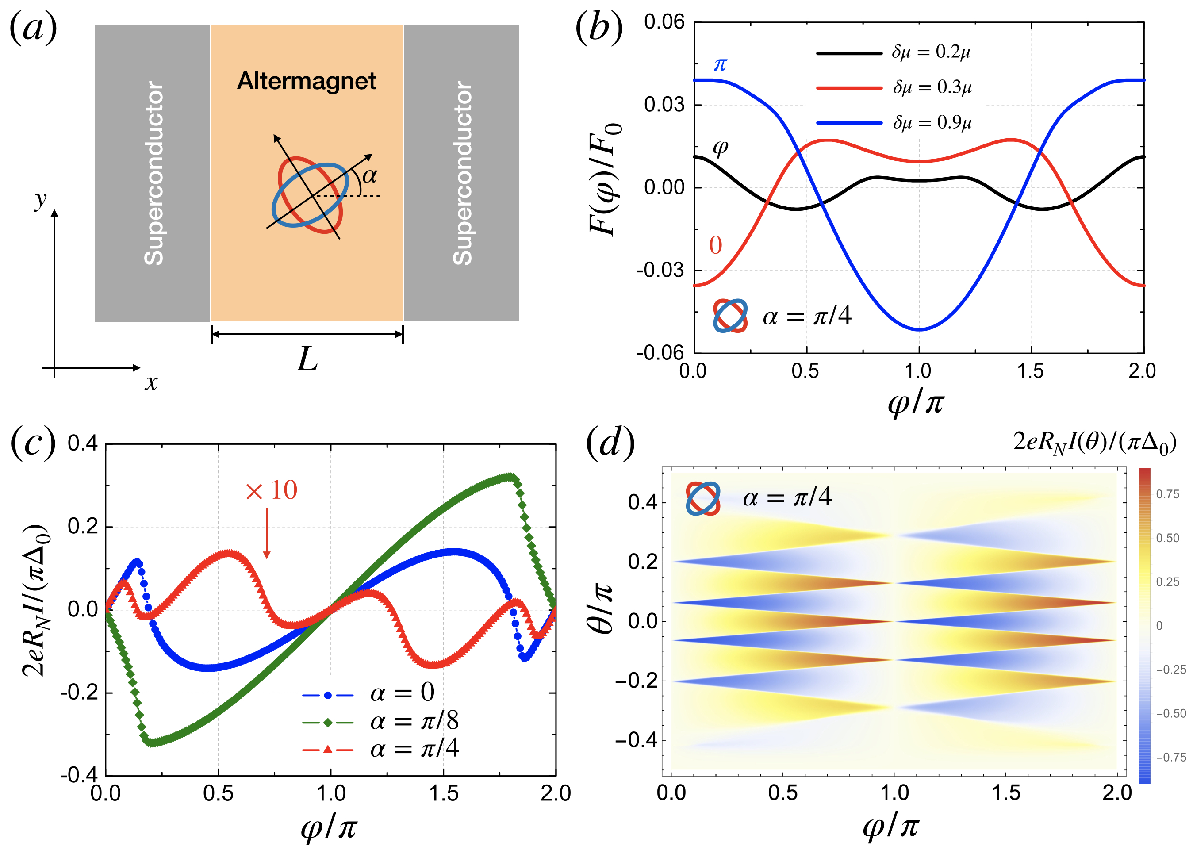}
	\end{center}
	\caption{(a) Schematic of the planar S/AM/S Josephson junction. (b) Free energy $F$ as a function of phase $\varphi$ 
     for various chemical potential differences $\delta \mu$ with a $d_{xy}$-AM ($\alpha=\pi/4$). The system exhibits transitions from $0$ (red), $\pi$ (blue) to $\varphi$ (black) junctions.  (c) The Josephson current for $\alpha=0$, $\alpha=\pi/8$ and $\alpha=\pi/4$ with $\delta \mu=0$. (d) The corresponding angle-resolved current for $\alpha=\pi/4$ in panel (c). $\theta$ is the injection angle $\theta=\arcsin{(k_y/k_F)}$. For panels (b)(c)(d), we have set $J/\mu$=0.2, $Z=0$, $k_F L=20$ and temperature $T=0.025T_c$.}
	\label{Fig1}
\end{figure}

In this letter, we study the Josephson effect of a two-dimensional {\color{blue} clean} S/AM/S junction, as depicted in Fig. \ref{Fig1}(a). We find that in striking contrast to the behavior of the Josephson current in conventional S/ferromagnet/S junctions, the S/AM/S junction can yield not only $0$ or $\pi$ junction, but also $\varphi$ junction via parameter change, see Fig. \ref{Fig1}(b). Anomalous CPR is also found with multiple nodal points (the red line in Fig. \ref{Fig1}(c)). 
These exotic results arise from the angle-resolved component of the Josephson current which sensitively depends on the momentum parallel to the interface. It is noted that the Cooper pairs induced in proximitized AM acquire a finite center-of-mass momentum in the $x$ direction \cite{zhang2024}. Such center-of-mass momentum varies for each $k_y$ channel because of the alternating spin-split field in the Brillouin zone. 
The angle-resolved components of Josephson current thus display a $0$-$\pi$ transition as shown in Fig.\ref{Fig1}(d), similar to the system with layered multiple $0$- and $\pi$-segments \cite{Pugach10}. The total current as the superposition of all angle-resolved components can have anomalous CPR with multiple nodal points, as well as $\varphi$ junction.
Our findings provide a way of realizing $\varphi$ junction without complicated geometry. 

\emph{Model and Formalism.---}
Our model consists of a $d$-wave altermagnet between two semi-infinite superconductors as shown in  Fig. \ref{Fig1}(a). We take the junction in $x$-direction where AM is located at $0<x<L$. In terms of the Nambu spinors $\hat{c}_{%
	\bm{k}}=(c_{\bm{k}\uparrow },c_{\bm{k}\downarrow },c_{-\bm{k}\uparrow
}^{\dag },c_{-\bm{k}\downarrow }^{\dag })^{T}$ with $\bm{k}=(k_x,k_y)$, the Hamiltonian of the
system is written as $\mathcal{H}=\mathrm{\frac{1}{2}}\sum\nolimits_{\bm{k}}\hat{c}_{%
	\bm{k}}^{\dag }\mathcal{H}_{\bm{k}}\hat{c}_{\bm{k}}$ with
\begin{equation}
	\mathcal{H}_{\bm{k}}=\left( \frac{\hbar ^{2}\bm{k}^{2}}{2m}-\tilde{\mu}+U\right) \hat{\tau}_{z}+M_{\bm{k}}\hat{s}_{z}\hat{\tau}_{z}-\tilde{\Delta} \hat{%
		s}_{y}\hat{\tau}_{y}.  \label{Hami}
\end{equation}
Here, $\tilde{\mu}(x)=\mu + \delta\mu \Theta(x)\Theta(L-x)$ is the chemical potential. $\delta\mu$ is a difference of chemical potential between S and AM which can be changeable by chemical doping.  $\hat{s}_{i}\left( \hat{\tau}%
_{i}\right) $ is the Pauli matrix in spin (Nambu) space. $M_{\bm{k}}$ denotes 
the altermagnetism and without loss of generality, the Neel vector of AM is along $z$-axis \cite{Libor011028}, 
\begin{equation}
	M_{\bm{k}}=\mathcal{J} k_F^{-2}\left[ \left( k_{x}^{2}-k_{y}^{2}\right) \cos 2\alpha
	+2k_{x}k_{y}\sin 2\alpha \right], 
\end{equation}
with $\mathcal{J}=J\Theta(x)\Theta(L-x)$ and $J$ the strength of the exchange energy of AM. $\alpha $ is the angle between the lobe of the direction of AM and normal to the interface, see Fig. \ref{Fig1} (a). We assume that $0<J<\mu /2$ to
have a well-defined Fermi surface \cite{Sun23}. $k_{F}$ is the wavevector $k_F=\sqrt{2m\mu/\hbar^2}$. For $\alpha =0$, the magnetization has
pure $d_{x^{2}-y^{2}}$-wave symmetry and for $\alpha =\pi /4$, it has pure $%
d_{xy}$-wave symmetry. We assume the conventional spin-singlet $s$-wave pair potential, 
the spatial dependence of which is given by $\tilde{\Delta} \left( x\right) =\Delta (T) \lbrack e^{i\varphi }\Theta \left(
-x\right) +\Theta \left( x-L\right) ]$ where $\varphi $ is the macroscopic
superconducting phase difference. We further adopt the BCS relation for its
temperature dependence: $\Delta \left( T\right) =\Delta _{0}\tanh (1.74\sqrt{%
	T_{c}/T-1})$ with $\Delta _{0}=1.76k_{B}T_{c}$, $T_{c}$ the critical
temperature, and temperature $T={k_B}^{-1}{\beta}^{-1}$. We have the barrier potential $U=U_{I}[\delta \left( x\right) +\delta \left( x-L\right) ]$, where $U_{I}$ is the fixed
barrier strength at the interface between S and AM and we define $Z=mU_I/(\hbar^2k_F)$.

\begin{figure}[t]
	\begin{center}
		\includegraphics[width=87mm]{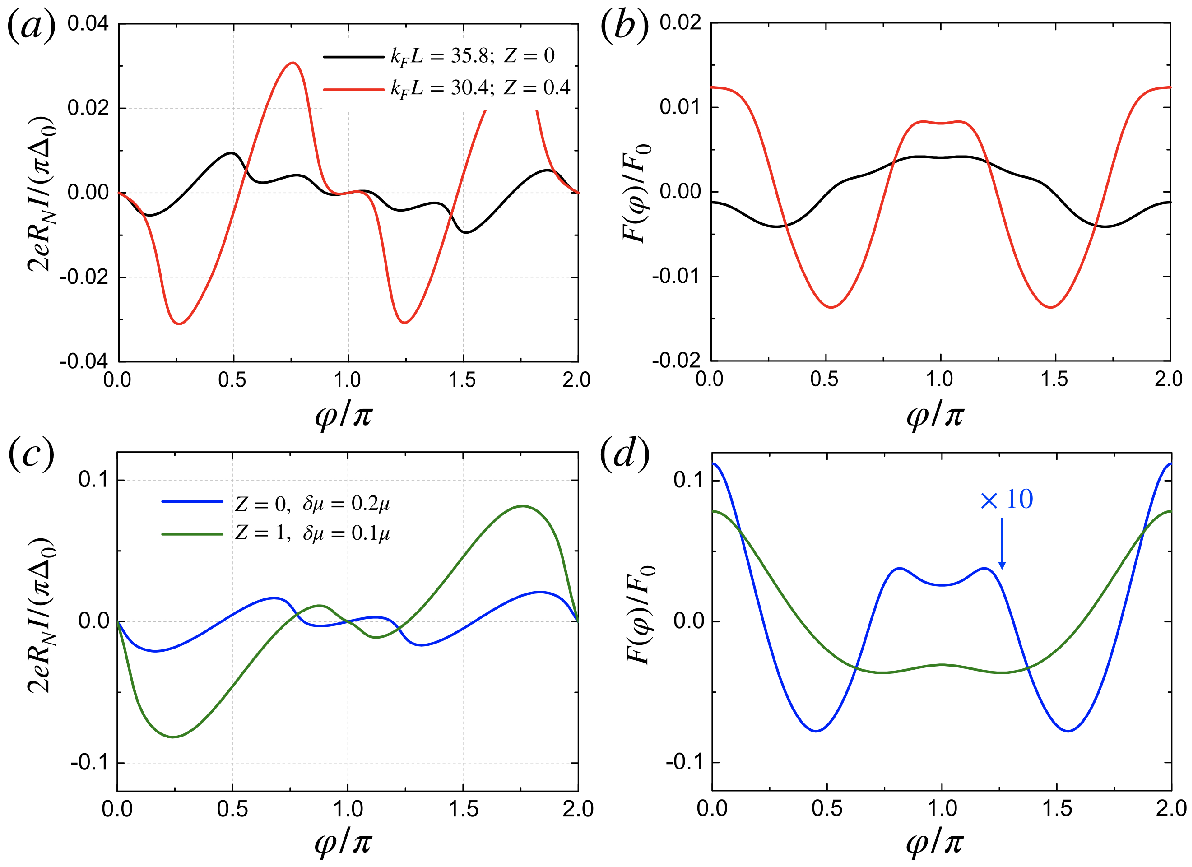}
	\end{center}
	\caption{Josephson currents and their free energies of $\varphi$ junction with a $d_{xy}$-AM: (a) Josephson current and (b) the corresponding free energy with equal chemical potential $\delta \mu=0$ for $Z=0$, $k_FL=35.8$ (black) and $Z=0.4$, $k_F L=30.4$ (red). (c) and (d) are Josephson current and free energy for fixed $k_FL=20$. The $\varphi$ junction is realized when $\delta \mu=0.2\mu$ with $Z=0$ (blue) and $\delta \mu=0.1\mu$ with $Z=1$ (green). Other parameters are $\alpha=\pi/4$, $J/\mu$=0.2 and $T=0.025T_c$. }
	\label{Fig2}
\end{figure}

\begin{figure*}[t]
	\begin{center}
		\includegraphics[width=165mm]{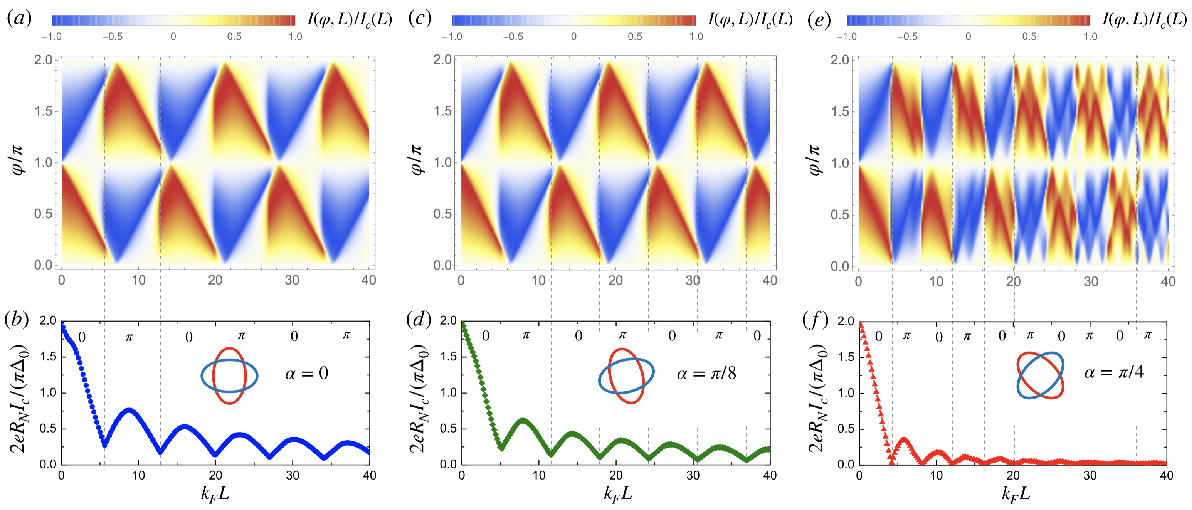}
	\end{center}
	\caption{The length dependence of Josephson current for different crystallographic orientation of AM. Upper panels: current phase relation $I(\varphi)$ as a function of the junction length for (a) $\alpha=0$, (c) $\pi/8$ and (e) $\pi/4$. For a specific $L$, the current is normalized to the maximum value $I_c(L)=\max ({I(\varphi,L)})$. Lower panels: the maximum Josephson current corresponding to the upper ones: (b) $\alpha=0$, (d) $\pi/8$ and (f) $\pi/4$. Other parameters are $J/\mu=0.2$, $\delta \mu=0$, $Z=0$. Temperature is at $T=0.025T_c$.}
	\label{Fig3}
\end{figure*}

The equilibrium Josephson current can be calculated by Furusaki-Tsukada 
formula \cite{FT91,Kashiwaya_2000} which includes contribution from both continuum and Andreev bound states. 
Due to the translational invariance along $y$-axis, the $y$-component
of the wave vector $k_{y}=\sqrt{2m\mu /\hbar^2}\sin \theta=k_F\sin\theta  $ is preserved,
where $\theta $ is the injection angle from $-\pi /2$ to $\pi /2$. Thus the system can be regarded as multiple transverse transport channels. 
We consider a very wide sample $W\gg L$ where $W$ is the junction width. So the summation over transverse momenta $k_{y}$ can be replaced by an
integration over $\theta$. For a
fixed $\theta $, we consider the four types of local Andreev reflection
coefficients by quasiparticle injections from the left S: a spin-$\uparrow $(%
$\downarrow $) electron to a spin-$\downarrow $($\uparrow $) hole: $%
a_{e,\uparrow \left( \downarrow \right) }$ and a spin-$\uparrow $($%
\downarrow $) hole to a spin-$\downarrow $($\uparrow $) electron: $%
a_{h,\uparrow \left( \downarrow \right) }$. The coefficients $%
a_{e,\uparrow \left( \downarrow \right) }$, $%
a_{h,\uparrow \left( \downarrow \right) }$ are obtained by boundary conditions \cite{Sun23, Papaj23} without assuming a universal tunnel probability for all transverse modes \cite{Beenakker23}, see Supplementary Materials (SMs). 
Then, the Josephson current is calculated by $I=\int_{-\pi /2}^{\pi
	/2}I\left( \theta \right) d\theta $ with
\begin{equation}
	I\left( \theta \right) =\frac{e\Delta (T) \cos \theta }{2\hbar \beta }%
	\sum\limits_{\omega _{n},s}\frac{k_{nx}^{+}+k_{nx}^{-}}{\sqrt{\omega
			_{n}^{2}+\Delta (T) ^{2}}}\left[ \frac{a_{e,s}}{k_{nx}^{+}}-\frac{a_{h,s}}{%
		k_{nx}^{-}}\right].
\end{equation}
Here, $k_{nx}^{\pm }=\sqrt{\frac{2m}{\hbar ^{2}}\left( \mu \pm i\sqrt{\omega
		_{n}^{2}+\Delta (T)^{2}}\right) -k_{y}^{2}}$ are the wavevectors of S and we have made analytical continuation of incident quasiparticle energy $ E \rightarrow i \omega_n$ into Matsubara frequencies $\omega _{n}=\pi k_{B}T(2n+1),(n=0,\pm
1,\pm 2....)$.
We consider the short junction $L\ll \xi $ where $\xi =\hbar
v_{F}/\Delta_0$ is the coherence length and choose $\Delta_0/\mu=0.01$.  The Fermi wavevector can be estimated as $k_F\approx 10^{10}$ $\text{m}^{-1}$ so that the typical junction length would be several nm in our model. We finally normalize $I$ to $2eR_N I / (\pi \Delta_0)$ where $R_N$ is the resistance of the junction in normal state, i.e., of the normal metal (N)/AM/N junction.

\emph{Symmetry analysis---} 
Before showing the numerical results, we analyze the general characteristic of CPR
from the symmetry point of view. We can decompose the Josephson current into
a series of different orders of Josephson coupling%
\begin{equation}
	I\left( \varphi \right) =\sum\nolimits_{n} [I_{n}\sin (n\varphi )+J_{n}\cos
	(n\varphi )].
\end{equation}%
We consider the fourfold rotation operator $C_{4} $ corresponding to a rotation angle $\pi /2$ with respect to $z$-axis which makes $k_{x}\rightarrow
k_{y}$, $k_{y}\rightarrow -k_{x}$, $\hat{s}_{z}\rightarrow \hat{s}_{z}$.
The altermagnetism flip its sign under $C_{4} $.  Using the time reversal operation $\mathcal{T}%
=-i\hat{s}_{y}\mathcal{K}$ with $\mathcal{K}$ the complex conjugation
operator, we can make $k_{x}\rightarrow -k_{x}$, $k_{y}\rightarrow -k_{y}$,  $\hat{s}_{z}\rightarrow -\hat{s}_{z}$ and $%
\varphi \rightarrow -\varphi $. Thus a combined operation with $\mathcal{T}$ i.e., $\mathcal{\tilde{M}%
	=T}C_{4} $ will give rise to
\begin{equation}
\mathcal{\tilde{M}}\mathcal{H}\left( \varphi \right) \mathcal{\tilde{M}}^{-1}=\mathcal{H}\left(
	-\varphi \right).
\end{equation}
Thus, the energy spectrum $E$ and the free energy $F$ is an
even function of $\varphi $: $F\left( \varphi \right) =F\left( -\varphi
\right) $ while the Josephson current $I\left( \varphi \right) \propto
\partial _{\varphi }F\left( \varphi \right) $ has the property $I\left(
\varphi \right) =-I\left( -\varphi \right) $ for arbitrary $\alpha$. It prohibits the term $J_{n}\cos(n\varphi )$ in the CPR of our system in the absence of spin-orbit couplings \cite{lu2015anomalous}. We will point out that although the characteristic of CPR is conventional but high-harmonic term $I_n \sin (n\varphi )$ ($n>1$) can become dominant in the presence of altermagnetism.  

\begin{figure*}[t]
	\begin{center}
		\includegraphics[width=165mm]{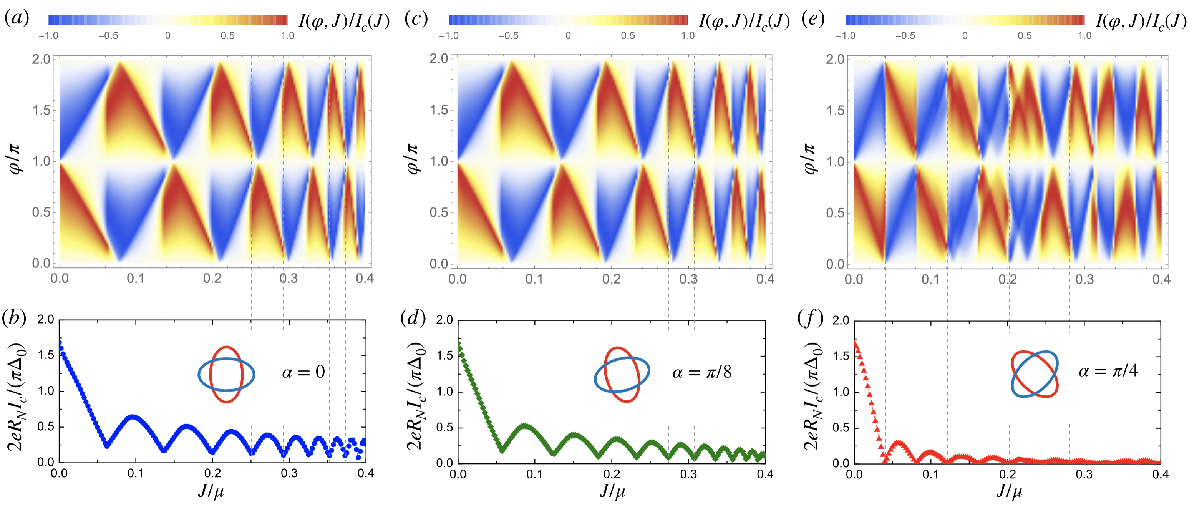}
	\end{center}
	\caption{The altermagnetic field dependence of Josephson current. Upper panels show the current phase relation as a function of the AM strength $J$ while the lower ones are the corresponding maximum supercurrent. $J$ yields $0$-$\pi$ transitions and decaying maximum current $I_c$. The crystallographic orientations of AM are $\alpha=0$ in (a)(b), $\pi/8$ in (c)(d) and $\pi/4$ in (e)(f).  We set $k_F L=20$, $\delta \mu=0$, $Z=0$ and temperature $T=0.025T_c$.}
	\label{Fig4}
\end{figure*}

\emph{Anomalous CPR---} 
We calculate the current-phase relation of the Josephson junction in \cref{Fig1}(c), with $k_F L=20$ ($L=0.1\xi$), the temperature $T=0.025T_{c}$ and $\delta \mu =0$.
\cref{Fig1}(c) shows the orientation-dependent CPRs for a fixed altermagnetic
strength $J=0.2\mu$. It is seen that the CPRs strongly depend on the
crystallographic orientation of AM. For $d_{x^{2}-y^{2}}$-AM ($\alpha =0$),
the CPR\ exhibits 4 nodes in one period $\varphi \in \lbrack 0,2\pi )$ while
the number of nodes becomes 8 for $d_{xy}$-AM ($\alpha =\pi /4$). When the
orientation angle $\alpha $ is $\pi /8$, quite
standard sinusoidal CPR\ is reproduced. 

To elucidate the origin of exotic
CPRs, we plot the angle-resolved Josephson current shown in \cref{Fig1}(d) which corresponds to a $d_{xy}$-AM. Clearly, for a fixed $\theta$, the direction of the current becomes either positive or negative depending on the angle. Except the nodal points at $\varphi=0$ or $\pi$ due to $I(\varphi)=-I(-\varphi)$ with $2\pi$ periodicity, the angular-averaged current as a function of $\theta$ can also have nodes when positive and negative contributions happen to cancel each other. The present multi-nodal CPR is a specific feature of the $d$-wave altermagnetism, where both the magnitude and the sign of the exchange field vary by changing $\theta$. The system corresponds to the multi-layered $0$ and $\pi$ S/ferromagnet/S junction in the momentum space, while each layer can develop the CPR independently. In this sense, the $\theta$ dependence of CPRs can generate Josephson current with multiple nodes for $d$-wave altermagnetism.

\emph{Tunable $\varphi$ Josephson junction---} Thus far we have shown that $0$-$\pi$ oscillation in the momentum space. Next, we explore the existence of $\varphi$ junction. We focus on the $d_{xy}$-AM since the angle-resolved CPR has the most drastic oscillations. We choose the same magnitude of $J$ as shown in \cref{Fig1} but change the junction length $L$. We find that the CPR is sensitive to the variation of $L$ and the $\varphi$ junction emerges. As an example, we plot the obtained $\varphi$ junctions for different values of $L$ in \cref{Fig2}(a).  We show the corresponding free energy $F(\varphi)$ in \cref{Fig2}(b), where $F$ is given by $I=2e \hbar^{-1} \partial F / \partial \varphi$.
We normalized the value of $F$ by $F_{0}=4e^{2}R_{N} / (\pi \hbar \Delta _{0})$. We have verified that the minimum value of free energy is located at two degenerate $\varphi$ ($ \neq 0$ or $\pi$). 

To see the feasibility that $\varphi$ junction can be widely formed and tuned in this setup, we consider the chemical potential difference between S and AM which can be controlled by chemical doping in AM, like copper oxides \cite{copper08}. It is shown that the variation of $\delta \mu$ can generate $\varphi$ junction in \cref{Fig1}(b) for a flat interface $Z=0$. In Figs.~\ref{Fig2}(c)(d), we see that it is also possible to obtain a $\varphi$ junction for finite interface barrier strength. In the SMs, we provide further examples of robust $\varphi$ junctions against non-ideal boundary conditions.  
Our result can be explained intuitively that the wavevector of AM for each $k_y$ varies by tuning $\delta \mu$, due to the nature of momentum-dependent polarization of AM where $M_{\bm{k}}$ has both the magnitude and sign change. Since the momentum-resolved CPRs rely sensitively on $M_{\bm{k}}$, which produces $0$-$\pi$ transition, the configuration of momentum-resolved CPRs as well as the total Josephson current by summing all channels can easily be changed by chemical doping. The merit of this tunability provides the control of CPR and realization of $\varphi$ junction in our geometry. 

\emph{$0$-$\pi$ oscillation---} We have given several examples of $\varphi$
junction for some certain parameters. Thus the S/AM/S system can yield all
three types of CPR, $0$, $\pi$, and $\varphi$. Next, we will explore the
condition for the transition of CPR by changing junction parameters and
discuss the condition for the emergence of $\varphi$ junction.

In \cref{Fig3}, we show the Josephson current as a function of $L$. The current $%
I(\varphi, L)$ has been normalized to the critical current with the same
length $I_{c}(L)$. It is shown in Figs. \ref{Fig3} (a)(c)(e) that $0$-$\pi $
transition occurs by changing $L$, in which a dip in the critical current $I_{c}$ develops, see %
Figs. \ref{Fig3} (b)(d)(f). Compared to the lattice model \cite{Ouassou23} where
the width along the $y$-axis is limited, our model incorporates more
channels in the large width limit along the $y$-axis. Thus, we obtain the
smooth periodic behavior of $I_{c}$. Moreover, the period of this
oscillation is longer for $d_{x^{2}-y^{2}}$-AM junction, see \cref{Fig3}(b),
but shorter for $d_{xy}$-AM one, see \cref{Fig3}(f). From the CPR, we find
that the high-harmonic components appear more frequently in $d_{xy}$-AM
junction for most values of $L$, as seen in \cref{Fig3}(e). The existence of
many high-harmonic components reduces the magnitude of $I_{c}$ and gives
rise to exotic CPRs. It can be explained that the positive and negative
values of angle-resolved Josephson current likely cancel each other near the
$0$-$\pi $ transition point. Since the momentum-resolved CPR varies
asynchronously, the slight change of various parameters in the model can
generate neither $0$ nor $\pi $ junction. On the other hand, in
S/ferromagnet/S junctions where momentum-resolved CPR behaves almost
synchronously close to the $0$-$\pi $ transition point for varying
parameters, the system becomes either $0$ or $\pi $ junction. We thus
conclude that the anomalous $\varphi $ dependence of the Josephson current
can occur in an AM junction near the $0$-$\pi $ transition point due to the
alternating magnetic orderings in the momentum space.

In \cref{Fig4}, we show the Josephson current as a function of $J$ while
keeping other parameters the same as \cref{Fig1}. We find that the variation
of $J$ can also induce $0$-$\pi$ transition in CPR as shown in Figs. \ref{Fig4}
(a)(c)(e) and oscillation in $I_{c}$, see Figs. \ref{Fig4} (b)(d)(f). $I_{c}$
decreases rapidly as $J$ grows, revealing that the Andreev reflection is suppressed
due to the broken time-reversal symmetry by the AM orderings. It is also
found that as $J$ grows, $0$-$\pi $ transition occurs more frequently. For the $%
d_{xy}$-AM junction in \cref{Fig4}(e), the CPR deviates from the
conventional sinusoidal one near the $0$-$\pi $ transition. From this
result, we can conclude that the CPR depends sensitively on the magnitude of
the exchange coupling of AM.

\emph{Conclusions.---} To summarize, we have studied the Josephson effect in a superconductor/altermagnet/superconductor junction.
It is found that anomalous phenomena including $0$-$\pi$ transition as well as multi-nodal current-phase relations appear. We can design $\varphi$ junction by tuning the exchange coupling, the orientation of crystal axis, the length, and the chemical doping of altermagnet.  These rich features serve as a guide to designing quantum two-level systems with large tunability in Josephson systems \cite{Ioffe1999}. In this letter, we focus on conventional spin-singlet $s$-wave 
superconductors. It is of more interest to study the altermagnetic Josephson junctions with unconventional superconductors, such as topological superconductors \cite{Qi,Tanaka12,Alicea2012,Beenakker2013,SatoAndo2017,Tanaka2024} in the future.



{\itshape Acknowledgment.} We thank S. Kashiwaya, S. Ikegaya, J. Cayao and Y. Fukaya for fruitful discussion. B. L. is supported by the National Natural
Science Foundation of China (project 11904257).
Y. T. acknowledges financial support from JSPS with Grants-in-Aid
for Scientific Research (KAKENHI Grants Nos. 23K17668, 24K00583, and 24K00556.).

\bibliography{altermagnet}

\end{document}